\documentclass[a4paper]{article}
\usepackage{hyperref}
\usepackage[dvipsnames]{xcolor}
\usepackage{multirow}
\usepackage{dblfloatfix}

\usepackage{INTERSPEECH2019}

\title{AdaDurIAN: Few-shot Adaptation for Neural Text-to-Speech with DurIAN}
\name{Zewang Zhang, Qiao Tian, Heng Lu, Ling-Hui Chen, Shan Liu}
\address{
  Tencent, China}
\email{\{zewangzhang, briantian, bearlu, nedchen, shiningliu\}@tencent.com}

\begin{document}

\maketitle
\begin{abstract}
This paper investigates how to leverage a DurIAN-based average model to enable a new speaker to have both accurate pronunciation and fluent cross-lingual speaking with very limited monolingual data. A weakness of the recently proposed end-to-end text-to-speech (TTS) systems is that robust alignment is hard to achieve, which hinders it to scale well with very limited data. To cope with this issue, we introduce AdaDurIAN by training an improved DurIAN-based average model and leverage it to few-shot learning with the shared speaker-independent content encoder across different speakers. Several few-shot learning tasks in our experiments show AdaDurIAN can outperform the baseline end-to-end system by a large margin. Subjective evaluations also show that AdaDurIAN yields higher mean opinion score (MOS) of naturalness and more preferences of speaker similarity. In addition, we also apply AdaDurIAN to emotion transfer tasks and demonstrate its promising performance.

\end{abstract}

\section{Introduction}
The rise of deep learning~\cite{lecun2015deep} has made more complex sequence generation tasks~\cite{sutskever2014sequence,Wang2017TacotronTE,shen2018natural,oord2016wavenet,Kalchbrenner2018EfficientNA} feasible. Text-based generation of natural speech has been continuously investigated over the past decades. Concatenative synthesis with unit selection~\cite{hunt1996unit} and statistical parametric speech synthesis~\cite{zen2009statistical} were the state-of-the-art systems for many years. However, such systems require lots of human labour and are unsatisfactory for lacking naturalness. Recently, a sequence-to-sequence architecture, Tacotron~\cite{Wang2017TacotronTE,shen2018natural}, has greatly improved the naturalness and similarity of speech synthesis compared to traditional statistical parametric speech synthesis system~\cite{zen2009statistical}. Tacotron, usually followed by a traditional or neural vocoder~\cite{oord2016wavenet,Griffin1984SignalEF}, takes linguistic feature and speaker identity as input and generates mel-spectrogram as output. Unfortunately, when dealing with out-of-domain or abnormal texts inputs, Tacotron-like attention based end-to-end structures could render unacceptable errors, including skipping, repeating, long unexpected pause and attention collapse~\cite{shen2018natural,tiantencent}. More recently, stepwise monotonic attention (SMA)~\cite{He2019RobustSA} method, which is based on monotonic attention~\cite{raffel2017online}, was proposed to enforce strict constraint to meet the demand of locality, monotonicity and completeness in the speech synthesis process. 

As far as we know, building a naturally speaking TTS system requires at least ten hours of recording audio. Moreover, every audio utterance should be recorded in a professional recording studio and the transcribed phonemes should be evenly distributed. Preparing such a large amount of high-quality data with multiple speakers is impractical and extremely expensive. Typically, it's troublesome and unnecessary to let native Chinese speaker to say English if he knows little about English. Moreover, there is no chance to gather 10 hours training data for a specific person like a pop star. The only resources we can get are the limited talks or shows from TV. Therefore, utilizing a few minutes of audio and synthesizing arbitrary speech in target's voice remains a very important task. 

However, building TTS system with limited data often sacrifices quality and reliability~\cite{chung2019semi}. To scale the capacity for new speakers, we can adapt existing pre-trained multi-speaker system to generate new speakers' voice, which is a well-studied subject of few-shot learning~\cite{fink2005object,fei2006one} also known as speaker adaptation~\cite{yamagishi2009analysis,leggetter1995maximum}. There are mainly two approaches here: the first is just to update the new speaker embedding and combine it with linguistic feature as inputs to a TTS model~\cite{jia2018transfer,li2017deep}, which may require a very strong speaker encoder network trained by thousands of speakers~\cite{8462665}; the second is to fine-tune the entire multi-speaker network to select a optimal single-speaker model~\cite{arik2018neural,chen2018sample,9054301}. Although fine-tuning can combine the advantages of multiple speakers and achieve a new speaker's better performance, as we described before, end-to-end attention models such as Tacotron-like models may meet unpredictable instability and bad cross-lingual speaking in few-shot learning settings. To achieve naturalness and robustness in speech synthesis, FastSpeech~\cite{ren2019fastspeech} and duration informed attention network (DurIAN)~\cite{Yu2019DurIANDI} have been recently proposed to overcome the unexpected errors of end-to-end systems by combining duration information of traditional statistical parametric speech synthesis system~\cite{zen2009statistical}. The former FastSpeech is a non-autoregressive feed-forward framework without attention. The latter DurIAN, originally proposed for multi-modal speech synthesis, is an autoregressive framework which achieves robustness and naturalness by using skip state encoder and combining duration with windowed content-based attention~\cite{Bahdanau2015NeuralMT}. 

\begin{figure}[htp]
    \centering
    \includegraphics[width=8cm]{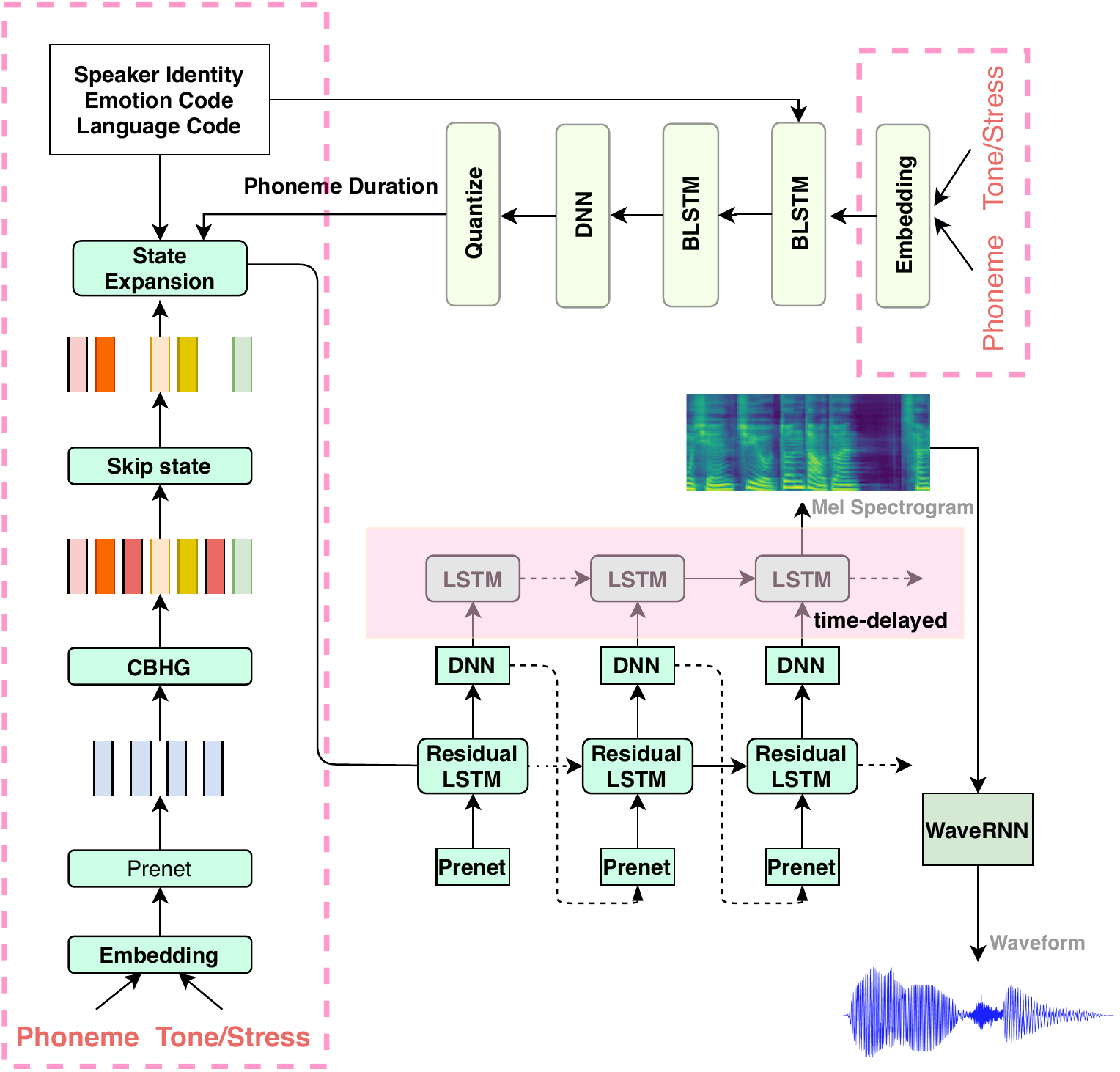}
    \caption{Detailed architecture of AdaDurIAN. Modules in the light red dotted rectangle will be fixed in few-shot speaker adaptation. For simplification, the windowed content-based attention is not shown.}
    \label{fig:AdaDurIANPro}
\end{figure}

To improve the scalability of TTS in few-shot speaker adaptation, we introduce AdaDurIAN, an adaptive neural TTS system based on DurIAN, with the ability to synthesize natural cross-lingual speech in a new speaker's voice with just few minutes of monolingual data. We investigate it in three different aspects that have not been fully explored in previous work. First, we employ sequences of phoneme and tone (or stress) to achieve a robust speaker-independent content encoder, and incorporate the concatenated representation of speaker characteristics into the output states of content encoder. Second, instead of fine-tuning weights of the whole architecture, we found a key aspect that only fine-tuning the speaker embedding and decoder network leads to fewer pronunciation errors. Last, to generate the smooth mel spectrograms in a streaming inference manner, we adopt a time-delayed LSTM post-net instead of a global CBHG-like~\cite{Wang2017TacotronTE} module. Through various evaluations, our proposed AdaDurIAN significantly surpasses the Tacotron-like model~\cite{He2019RobustSA} in terms of naturalness, speaker similarity and cross-lingual speaking, and also shows its promising performance in few-shot emotion transfer tasks.

The rest of this paper is organized as follows. Section~\ref{section2} describes the detailed architecture of AdaDurIAN and the speaker adaptation strategy. The experiment setup and evaluations are presented in Section $3$. Concluding remarks are summarized in the final section.

\section{The proposed method}
\label{section2}
\subsection{Architecture of AdaDurIAN}

%% DurIAN~\cite{Yu2019DurIANDI} was originally proposed for multi-modal speech synthesis, which achieves both robustness and stability without losing naturalness by replacing the end-to-end attention mechanism in the Tacotron 2~\cite{shen2018natural} with duration model in the traditional parametric speech synthesis system~\cite{zen2009statistical}. Besides, a skip encoder architecture is used for encoding both the phoneme sequence and hierarchical prosodic structure for improved generalization in Chinese speech synthesis tasks. 

The original DurIAN~\cite{Yu2019DurIANDI} is a single-speaker TTS system, the model for each speaker should be trained individually with their own voice. We made improvements to support multi-speaker, multi-style and multi-lingual speech synthesis. Figure~\ref{fig:AdaDurIANPro} shows the architecture of proposed AdaDurIAN. It's composed of (1) a speaker-independent content encoder that encodes the linguistic sequences, (2) an alignment model that predicts the duration of each phoneme and then aligns the output states of content encoder to acoustic frames, (3) a decoder network that generates $5$ frames of mel spectrogram autoregressively.

\subsubsection{Speaker-independent Content Encoder}
It's hard to ensure that the input tokens (phonemes, tones, stresses and so on) are evenly distributed for a single speaker's training corpus. To benefit from the knowledge of multi-speaker's training corpus, different from DurIAN, we take both phoneme and tone (or stress, which appears in English words) sequence with prosodic boundary symbols as the input to content encoder of AdaDurIAN. With state skipping~\cite{Yu2019DurIANDI}, the output of the content encoder is a sequence of hidden states containing speaker-independent global linguistic feature transformation.

\subsubsection{Alignment Model}
To combine linguistic feature transformation and speaker characteristics representation, we incorporate speaker embedding, emotion embedding and language embedding into the expanded states of content encoder. The language code switching is implemented based on the language to which the current phoneme belongs. Such speaker-dependent concatenated representation makes AdaDurIAN to synthesize speech for different speakers with different styles. Each speaker-dependent frame state is repeated according to the alignment model and then concatenated with relative position encoding~\cite{Yu2019DurIANDI} inside each phoneme. The detailed structure of alignment model is shown in Figure~\ref{fig:AdaDurIANPro}, and the alignment model doesn't share any trainable embeddings with content encoder for stable training.

\subsubsection{Decoder}
Different from DurIAN, we adopt the residual LSTM~\cite{Kim2017ResidualLD} layers for its efficient training which is of great importance in few-shot learning. Instead of CBHG~\cite{Wang2017TacotronTE} module, we adopt a vanilla LSTM layer with a time delay of $5$ frames as post-net. Practically, such structure of post-net can significantly improve the quality of mel spectrogram predicted by the decoder, and also achieve the streaming synthesis to be deployed in production environment. As a result, the inference of AdaDurIAN is $17$ times faster than real-time with two CPU cores.

\subsection{Speaker Adaptation Strategy}
Few-shot speaker adaptation is an intractable task for that there are very few training samples. The insoluble dilemma lying in few-shot speaker adaptation is that the distribution of linguistic tokens is hardly even. Following the general training procedure, the model of new speaker would soon be unable to synthesize out-of-domain words, let alone the naturalness and speaker similarity. Fortunately, the speaker-independent content encoder of AdaDurIAN can absorb the knowledge across different speakers, so a pre-trained content encoder can also be borrowed to transform linguistic feature for any new speaker.

\begin{table}[]
\centering
\caption{Results of word error rate (WER) with different modules to be fixed during few-shot speaker adaptation.}
\begin{tabular}{l|l}
Freeze Modules         & WER    \\ \hline
nothing                & 5.48\% \\ \hline
+phone embedding       & 4.00\% \\
\hskip 1em +tone/stress,language embedding & 3.36\% \\
\hskip 2em +encoder               & \textbf{2.28\%} \\ \hline
\end{tabular}
\label{tab:wer}
\end{table}

\begin{table*}[htbp]
\centering
\caption{Robust mean opinion score (MOS) as a function of the amount of training data and different speakers. CN speaker denotes a native Chinese speaker, while EN speaker denotes a native English speaker. All audios are converted by Griffin-Lim algorithm~\cite{Griffin1984SignalEF} from mel spectrogtam.}

\begin{tabular}{lll|ll|c|l}
\hline
\multicolumn{1}{c}{\multirow{2}{*}{\textbf{System}}} & \multicolumn{2}{c}{\textbf{Chinese sentences}}                & \multicolumn{2}{c}{\textbf{English sentences}}               & \multicolumn{2}{c}{\textbf{Recordings}}                              \\ \cline{2-7}
\multicolumn{1}{c}{}                        & \multicolumn{1}{l}{CN speaker} & EN speaker         & \multicolumn{1}{l}{CN speaker} & EN speaker         & \multicolumn{1}{l}{CN speaker} & EN speaker                 \\ \hline
\textbf{SMA} 1m  & $3.85\pm0.07$ & $3.29\pm0.07$ & $3.80\pm0.20$  & $3.55\pm0.21$ & \multirow{6}{*}{$4.45\pm0.05$} & \multirow{6}{*}{$4.14\pm0.06$} \\
\textbf{AdaDurIAN} 1m                                & $\mathbf{4.14\pm0.06}$             & $\mathbf{3.90\pm0.06}$ & $\mathbf{4.13\pm0.22}$              & $\mathbf{3.98\pm0.18}$ &                                 &                            \\ \cline{1-5}
\textbf{SMA} 3m                                      & $3.84\pm0.07$                       & $3.29\pm0.08 $         & $3.69\pm0.23 $                      & $3.18\pm0.24 $         &                                 &                            \\
\textbf{AdaDurIAN} 3m                                & $\mathbf{4.21\pm0.05}$              & $\mathbf{3.73\pm0.07}$ & $\mathbf{4.03\pm0.21}$              & $\mathbf{3.95\pm0.18}$ &                                 &                            \\ \cline{1-5}
\textbf{SMA} 20m                                    & $4.10\pm0.06 $                      & $2.97\pm0.09$          & $4.06\pm0.18 $                      & $2.58\pm0.28  $        &                                 &                            \\
\textbf{AdaDurIAN} 20m                               & $\mathbf{4.19\pm0.06}$              & $\mathbf{3.57\pm0.07}$ & $\mathbf{4.21\pm0.16}$              & $\mathbf{3.96\pm0.16}$ &                                 &                            \\ \hline
\end{tabular}
\label{tab:mos}
\end{table*}

Straightforwardly, the training procedure of AdaDurIAN for few-shot speaker adaptation is to transfer the linguistic feature transformation of a multi-speaker system to a new speaker with limited training data without losing naturalness, speaker similarity and cross-lingual speaking. Inspired by~\cite{Fan2015MultispeakerMA}, modules in the light red dotted rectangle in Figure~\ref{fig:AdaDurIANPro} will be fixed and shared for any new speaker. To achieve this, we first fully shuffle the training data to ensure that each mini-batch contains the data of different speakers and then train AdaDurIAN to get an average multi-speaker TTS model. At the stage of few-shot speaker adaptation, modules including phone embedding, tone(or stress) embedding, language embedding, emotion embedding and encoder in average model will be fixed. With such proposed speaker adaptation strategy, AdaDurIAN can be applied to speakers who have very limited data. We will validate that, by borrowing knowledge from other speakers and only optimizing speaker embedding and decoder, AdaDurIAN has a better performance in terms of naturalness, speaker similarity and cross-lingual speaking.

\section{Experiments}

We take SMA~\cite{He2019RobustSA} as our baseline model, an optimal variant of Tacotron 2~\cite{shen2018natural}, in which the memory at each decoding step is computed by a stepwise monotonic attention~\cite{He2019RobustSA} instead of an alignment model. We find that, compared with original Tacotron 2, SMA performs much better in terms of accurate pronunciation and synthesizing long or abnormal utterances. 

We performed three sets of experiments for adaptive TTS to show the performance of the proposed AdaDurIAN system. First, we investigate the stability of pronunciation under different adaptation strategies on AdaDurIAN. Second, we compare the performances of SMA and AdaDurIAN with a random subset of the audio with total duration of $1$, $3$ and $20$ minutes, respectively. Finally, we perform the few-shot emotion transfer tasks on two unseen speakers with limited neutral speech data. We highly recommend readers to go listen to the generated audios~\footnote{\url{https://xusongvae.github.io/adadurian}}. %此处需要插入网页版链接

\subsection{Experiment Setup}
The data we used is our internal carefully annotated $200$-hour speech corpus which is collected from around $55$ speakers with different genders and nations. All audios are sampled by $24$kHz with mono channel, windowed with $45$ ms and shifted every $10$ ms. The $80$-th order mel spectrograms are extracted to represent the spectral envelope. 

Two neural TTS systems are implemented for comparison. For AdaDurIAN, as shown in Figure~\ref{fig:AdaDurIANPro}, sequences of linguistic tokens are passed through a pre-net that contains three fully-connected layers followed by a CBHG module. The same group of sequences is taken as the input to duration model composed of two BLSTM layers with $512$ units. The output of content encoder would be expanded according to the duration of each phoneme. The pre-net of decoder is composed of two $256$-unit fully-connected layers. The expanded state and output of decoder pre-net are passed through into a content-based $tanh$ attention with depth $512$. Finally, the output of attention is passed through the bottom residual LSTM layer. At each decoding step, the second LSTM layer generates $4$ non-overlapped frames of mel spectrogram. A post-net with two stacked fully-connected layers with $512$ and $256$ hidden units, followed by a $256$-unit LSTM layer with a time delay of $5$ frames, consequently makes the mel spectrogram more smooth and generated in a streaming manner. As for the baseline model SMA, the architecture of which used in this paper is almost the same as AdaDurIAN, except that the unique attention mechanism of SMA is implemented with a GRU cell of $256$ unit. 

We first trained two average models with $300$k steps for SMA and AdaDurIAN, and then conducted speaker adaptation and emotion transfer tasks with the already described strategy, respectively. The batch size is $2$ and the validation step interval is $10$ to get a better model. The procedure of gradient descent optimization is the same as the original DurIAN~\cite{Yu2019DurIANDI}. 

In addition to the acoustic model, we used the robust and fast WaveRNN~\cite{Kalchbrenner2018EfficientNA} variant as vocoder, which consists of $1$D convolution as condition network and sparse GRU, four fully-connected layers with dual softmax structure. We trained such WaveRNN in the previously described dataset collected from around $55$ speakers, excluding the speakers to be evaluated in this paper. To eliminate the influence of WaveRNN, in MOS test, audios including recording audios are all converted by Griffin-Lim algorithm~\cite{Griffin1984SignalEF}, while audios of other tests are converted by such speaker-independent WaveRNN.

\subsection{Objective Evaluation}
We performed pronunciation error statistical task by using $1$-minute data of several speakers to compare the performances of AdaDurIAN under different few-shot speaker adaptation strategies. We randomly selected $30$ long and abnormal sentences with a total of $4000$ words for synthesis. Such pronunciation error statistical task was performed with $50$-$80$ anonymous and untrained subjects participating in several evaluation sessions, constructed so that each sentence was evaluated by $10$ dinstinct subjects. Each participant was asked to count the number of errors in each sentence, including wrong pronunciation, unclearness and incorrect tone. Although we can use automatic speech recognition (ASR) system, we find that ASR system is too robust to spot minor pronunciation errors.

We evaluate the performance of each adaptation strategy by calculating the word error rate (WER). Table~\ref{tab:wer} shows each WER of different adaptation strategies. We find that fixing phone embedding, tone (or stress) embedding, language embedding and encoder could achieve the least pronunciation errors, which is reasonable because these fixed parameters still stay in the same distribution space even given very limited unbalanced data. With extremely imbalanced $1$-min data, a much lower WER indicates that such fine-tuning strategy is reliable in few-shot speaker adaptation.

\begin{table*}[htbp]
\centering
\caption{The ABX test of speaker similarity.}
\begin{tabular}{l|l|lll}
\hline
\multicolumn{1}{c|}{\multirow{2}{*}{\textbf{System}}} & \multicolumn{2}{c|}{\textbf{Chinese sentences}}                   & \multicolumn{2}{c}{\textbf{English sentences}}                    \\ \cline{2-5}
\multicolumn{1}{c|}{}                        & \textit{CN} speaker       & \multicolumn{1}{l|}{\textit{EN} speaker}       & \multicolumn{1}{l|}{\textit{CN} speaker}       & \textit{EN} speaker       \\ \hline
\textbf{SMA} 1m                                       & 17.78\%          & \multicolumn{1}{l|}{7.22\%}           & \multicolumn{1}{l|}{8.33\%}           & 20.00\%          \\
\textbf{AdaDurIAN} 1m                                 & \textbf{62.96\%} & \multicolumn{1}{l|}{46.30\%}          & \multicolumn{1}{l|}{\textbf{51.67\%}} & \textbf{53.33\%} \\
No preference                                         & 19.26\%          & \multicolumn{1}{l|}{\textbf{46.48\%}} & \multicolumn{1}{l|}{40.00\%}          & 26.67\%          \\ \hline
\textbf{SMA} 3m                                       & 16.85\%          & \multicolumn{1}{l|}{22.78\%}          & \multicolumn{1}{l|}{13.33\%}          & 20.00\%          \\
\textbf{AdaDurIAN} 3m                                 & \textbf{70.56\%} & \multicolumn{1}{l|}{\textbf{44.07\%}} & \multicolumn{1}{l|}{\textbf{61.67\%}} & \textbf{68.33\%} \\
No preference                                         & 12.59\%          & \multicolumn{1}{l|}{33.15\%}          & \multicolumn{1}{l|}{25.00\%}          & 11.67\%          \\ \hline
\textbf{SMA} 20m                                      & 23.89\%          & \multicolumn{1}{l|}{16.48\%}          & \multicolumn{1}{l|}{20.00\%}          & 16.67\%          \\
\textbf{AdaDurIAN} 20m                                & \textbf{56.85\%} & \multicolumn{1}{l|}{36.11\%}          & \multicolumn{1}{l|}{\textbf{65.00\%}} & \textbf{61.67\%} \\
No preference                                         & 19.26\%          & \multicolumn{1}{l|}{\textbf{47.41\%}} & \multicolumn{1}{l|}{15.00\%}          & 21.67\%          \\ \hline
\end{tabular}

\label{tab:abx}
\end{table*}

\subsection{Subjective Evaluation}
\subsubsection{Speaker Adaptation}

 We selected one native Chinese speaker (denoted as ``\textit{CN} speaker'') without English speech corpus, and one native English speaker (denoted as ``\textit{EN} speaker'') without Chinese speech corpus to perform few-shot speaker adaptation tasks. We constructed three datasets for each test speaker with a total duration of $1$ minute, $3$ minutes and $20$ minutes, respectively. There is a held-out validation set for each speaker. Each few-shot speaker adaptation system was trained and selected according to the lowest validation loss. Then, we synthesized $30$ Chinese sentences and English sentences that were excluded from all training samples. We performed two subjective tests on these sentences. In the first test, Mean Opinion Score (MOS) test, for each generated audio, the subjects were asked to rate each audio from lowest score $1$ to highest score $5$ on the naturalness. In the second test, the similarity ABX test, the subjects were asked to listen a recorded audio first, and then choose which of the converted audios sounds more like that recorded audio or neither.

As shown in Table~\ref{tab:mos}, for both Chinese sentences and English sentences, AdaDurIAN has a higher MOS of naturalness over SMA. Especially, \textit{CN} speaker has a MOS $4.21$ when given only $3$ minutes of training data and there is only a MOS gap of $0.2$ compared with MOS of recordings. Given just $1$ minute of data, \textit{CN} speaker with AdaDurIAN system can still get a MOS $4.14$ on Chinese sentences and $4.13$ on English sentences, respectively. Although \textit{EN} speaker with AdaDurIAN has a decreasing MOS with the increase of data on Chinese sentences, the highest MOS $3.90$ only has a minor distance to the MOS $4.14$ of the recordings. This indicates that AdaDurIAN can generalize well in cross-lingual speaking even if few minutes of monolingual data is given.

\begin{table}[h]
\centering
\caption{Emotion preference test of female-to-male (F2M) emotion transfer with $1$ minute of data.}
\begin{tabular}{*6c}
\toprule
\multirow{2}{*}{\textbf{Target}} &  \multicolumn{4}{c}{\textbf{Selected}} & \multirow{2}{*}{\textbf{Acc.}}\\\cline{2-5}

{}   & \textbf{neutral}   & \textbf{anger}    & \textbf{happiness}   & \textbf{sadness} &\\\hline
\textbf{neutral}   &  \textbf{224} & 9   & 53  & 14 & 0.75 \\
\textbf{anger}   &  80 & \textbf{168}   & 32  & 20 & 0.56 \\
\textbf{happiness}   &  90  &  33   & \textbf{169}  & 8 & 0.56 \\
\textbf{sadness}   &  194  &  26   & 21  & 59 & 0.20\\\hline
\textbf{Average}   &    &     &   &  & \textbf{0.52}\\
\bottomrule
\end{tabular}

\label{tab:f2m}
\end{table}

\begin{table}[h]
\centering
\caption{Emotion preference test of female-to-female (F2F) emotion transfer with $1$ minute of data.}
\begin{tabular}{*6c}
\toprule
\multirow{2}{*}{\textbf{Target}} &  \multicolumn{4}{c}{\textbf{Selected}} & \multirow{2}{*}{\textbf{Acc.}}\\\cline{2-5}

{}   & \textbf{neutral}   & \textbf{anger}    & \textbf{happiness}   & \textbf{sadness} &\\\hline
\textbf{neutral}   &  \textbf{232} & 7   & 53  & 8 & 0.77 \\
\textbf{anger}   &  72 & \textbf{168}   & 16  & 44 & 0.56 \\
\textbf{happiness}   &  83  &  5   & \textbf{212}  & 0 & 0.71 \\
\textbf{sadness}   &  118  &  15   & 20  & \textbf{147} & 0.49\\\hline
\textbf{Average}&    &     &   &  & \textbf{0.63}\\
\bottomrule
\end{tabular}

\label{tab:f2f}
\end{table}

The result of speaker similarity preference test is shown in Table~\ref{tab:abx}. For \textit{CN} speaker, AdaDurIAN gains much more preferences than SMA in all experiments. For \textit{EN} speaker, AdaDurIAN outperforms SMA with a significant margin on English sentences, and AdaDurIAN still has a comparable performance with SMA on Chinese sentences. Such promising evaluation results motivate us to apply AdaDurIAN in further tasks that have a high requirement of speaker similarity.

\subsubsection{Emotion Transfer}
To explore the ability of AdaDurIAN on transfering different emotions with few neutral data, we used an available female corpus with four annotated emotion styles: neutral, anger, happiness and sadness. We first trained such a base emotional model on the previous AdaDurIAN average model, then we fine-tuned such female emotional model with a $1$ minute male speech corpus and a $1$ minute female speech corpus with already described strategy. We evaluate the performances of two female-to-male (F2M) and female-to-female (F2F) few-shot emotion transfer tasks by subjective emotion classification. As shown in Table~\ref{tab:f2m} and Table~\ref{tab:f2f}, emotion transfer of F2F task is less difficult than that of F2M. The mean emotion classification accuracy of F2F is $64\%$ while that of F2M is only $52\%$. Specifically, we find that the emotion transfer of neutral and happiness is the easiest, emotion transfer of sadness is the second, while emotion transfer of anger is the hardest. This discovery provides an important reference for future few-shot emotion transfer research.

\section{Conclusions}
In summary, we proposed AdaDurIAN, a few-shot adaptive neural TTS system for higher naturalness and speaker similarity. We described the improvements of AdaDurIAN over original DurIAN and demonstrated the adaptation strategy when the speaker's data is very limited. Based on AdaDurIAN, we performed several few-shot speaker adaptation tasks to evaluate the stability, naturalness, speaker similarity and emotion transfer ability. The evaluations show that, compared with Tacotron-like model, AdaDurIAN has both higher MOS of naturalness, more preferences of speaker similarity and especially fluent cross-lingual speaking. Furthermore, we also applied AdaDurIAN in emotion transfer tasks and showed its promising performance.

\vfill\pagebreak
\bibliographystyle{IEEEtran}

\bibliography{mybib}

\begin{thebibliography}{10}
\providecommand{\url}[1]{#1}
\csname url@samestyle\endcsname
\providecommand{\newblock}{\relax}
\providecommand{\bibinfo}[2]{#2}
\providecommand{\BIBentrySTDinterwordspacing}{\spaceskip=0pt\relax}
\providecommand{\BIBentryALTinterwordstretchfactor}{4}
\providecommand{\BIBentryALTinterwordspacing}{\spaceskip=\fontdimen2\font plus
\BIBentryALTinterwordstretchfactor\fontdimen3\font minus
  \fontdimen4\font\relax}
\providecommand{\BIBforeignlanguage}[2]{{%
\expandafter\ifx\csname l@#1\endcsname\relax
\typeout{** WARNING: IEEEtran.bst: No hyphenation pattern has been}%
\typeout{** loaded for the language `#1'. Using the pattern for}%
\typeout{** the default language instead.}%
\else
\language=\csname l@#1\endcsname
\fi
#2}}
\providecommand{\BIBdecl}{\relax}
\BIBdecl

\bibitem{lecun2015deep}
Y.~LeCun, Y.~Bengio, and G.~Hinton, ``Deep learning,'' \emph{nature}, vol. 521,
  no. 7553, pp. 436--444, 2015.

\bibitem{sutskever2014sequence}
I.~Sutskever, O.~Vinyals, and Q.~V. Le, ``Sequence to sequence learning with
  neural networks,'' in \emph{Advances in neural information processing
  systems}, 2014, pp. 3104--3112.

\bibitem{Wang2017TacotronTE}
Y.~Wang, R.~J. Skerry-Ryan, D.~Stanton, Y.~Wu, R.~J. Weiss, N.~Jaitly, Z.~Yang,
  Y.~Xiao, Z.~Chen, S.~Bengio, Q.~V. Le, Y.~Agiomyrgiannakis, R.~Clark, and
  R.~A. Saurous, ``Tacotron: Towards end-to-end speech synthesis,'' in
  \emph{INTERSPEECH}, 2017.

\bibitem{shen2018natural}
J.~Shen, R.~Pang, R.~J. Weiss, M.~Schuster, N.~Jaitly, Z.~Yang, Z.~Chen,
  Y.~Zhang, Y.~Wang, R.~Skerrv-Ryan \emph{et~al.}, ``Natural tts synthesis by
  conditioning wavenet on mel spectrogram predictions,'' in \emph{2018 IEEE
  International Conference on Acoustics, Speech and Signal Processing
  (ICASSP)}.\hskip 1em plus 0.5em minus 0.4em\relax IEEE, 2018, pp. 4779--4783.

\bibitem{oord2016wavenet}
A.~v.~d. Oord, S.~Dieleman, H.~Zen, K.~Simonyan, O.~Vinyals, A.~Graves,
  N.~Kalchbrenner, A.~Senior, and K.~Kavukcuoglu, ``Wavenet: A generative model
  for raw audio,'' \emph{arXiv preprint arXiv:1609.03499}, 2016.

\bibitem{Kalchbrenner2018EfficientNA}
N.~Kalchbrenner, E.~Elsen, K.~Simonyan, S.~Noury, N.~Casagrande, E.~Lockhart,
  F.~Stimberg, A.~van~den Oord, S.~Dieleman, and K.~Kavukcuoglu, ``Efficient
  neural audio synthesis,'' in \emph{ICML}, 2018.

\bibitem{hunt1996unit}
A.~J. Hunt and A.~W. Black, ``Unit selection in a concatenative speech
  synthesis system using a large speech database,'' in \emph{1996 IEEE
  International Conference on Acoustics, Speech, and Signal Processing
  Conference Proceedings}, vol.~1.\hskip 1em plus 0.5em minus 0.4em\relax IEEE,
  1996, pp. 373--376.

\bibitem{zen2009statistical}
H.~Zen, K.~Tokuda, and A.~W. Black, ``Statistical parametric speech
  synthesis,'' \emph{speech communication}, vol.~51, no.~11, pp. 1039--1064,
  2009.

\bibitem{Griffin1984SignalEF}
D.~W. Griffin and J.~S. Lim, ``Signal estimation from modified short-time
  fourier transform,'' \emph{IEEE Transactions on Acoustics, Speech, and Signal
  Processing}, vol.~32, pp. 236--243, 1984.

\bibitem{He2019RobustSA}
M.~He, Y.~Deng, and L.~He, ``Robust sequence-to-sequence acoustic modeling with
  stepwise monotonic attention for neural tts,'' in \emph{INTERSPEECH}, 2019.

\bibitem{raffel2017online}
C.~Raffel, M.-T. Luong, P.~J. Liu, R.~J. Weiss, and D.~Eck, ``Online and
  linear-time attention by enforcing monotonic alignments,'' in
  \emph{Proceedings of the 34th International Conference on Machine
  Learning-Volume 70}.\hskip 1em plus 0.5em minus 0.4em\relax JMLR. org, 2017,
  pp. 2837--2846.

\bibitem{chung2019semi}
Y.-A. Chung, Y.~Wang, W.-N. Hsu, Y.~Zhang, and R.~Skerry-Ryan,
  ``Semi-supervised training for improving data efficiency in end-to-end speech
  synthesis,'' in \emph{ICASSP 2019-2019 IEEE International Conference on
  Acoustics, Speech and Signal Processing (ICASSP)}.\hskip 1em plus 0.5em minus
  0.4em\relax IEEE, 2019, pp. 6940--6944.

\bibitem{fink2005object}
M.~Fink, ``Object classification from a single example utilizing class
  relevance metrics,'' in \emph{Advances in neural information processing
  systems}, 2005, pp. 449--456.

\bibitem{fei2006one}
L.~Fei-Fei, R.~Fergus, and P.~Perona, ``One-shot learning of object
  categories,'' \emph{IEEE transactions on pattern analysis and machine
  intelligence}, vol.~28, no.~4, pp. 594--611, 2006.

\bibitem{yamagishi2009analysis}
J.~Yamagishi, T.~Kobayashi, Y.~Nakano, K.~Ogata, and J.~Isogai, ``Analysis of
  speaker adaptation algorithms for hmm-based speech synthesis and a
  constrained smaplr adaptation algorithm,'' \emph{IEEE Transactions on Audio,
  Speech, and Language Processing}, vol.~17, no.~1, pp. 66--83, 2009.

\bibitem{leggetter1995maximum}
C.~J. Leggetter and P.~C. Woodland, ``Maximum likelihood linear regression for
  speaker adaptation of continuous density hidden markov models,''
  \emph{Computer speech \& language}, vol.~9, no.~2, pp. 171--185, 1995.

\bibitem{jia2018transfer}
Y.~Jia, Y.~Zhang, R.~Weiss, Q.~Wang, J.~Shen, F.~Ren, P.~Nguyen, R.~Pang, I.~L.
  Moreno, Y.~Wu \emph{et~al.}, ``Transfer learning from speaker verification to
  multispeaker text-to-speech synthesis,'' in \emph{Advances in neural
  information processing systems}, 2018, pp. 4480--4490.

\bibitem{li2017deep}
C.~Li, X.~Ma, B.~Jiang, X.~Li, X.~Zhang, X.~Liu, Y.~Cao, A.~Kannan, and Z.~Zhu,
  ``Deep speaker: an end-to-end neural speaker embedding system,'' \emph{arXiv
  preprint arXiv:1705.02304}, 2017.

\bibitem{8462665}
L.~{Wan}, Q.~{Wang}, A.~{Papir}, and I.~L. {Moreno}, ``Generalized end-to-end
  loss for speaker verification,'' in \emph{2018 IEEE International Conference
  on Acoustics, Speech and Signal Processing (ICASSP)}, 2018, pp. 4879--4883.

\bibitem{arik2018neural}
S.~Arik, J.~Chen, K.~Peng, W.~Ping, and Y.~Zhou, ``Neural voice cloning with a
  few samples,'' in \emph{Advances in Neural Information Processing Systems},
  2018, pp. 10\,019--10\,029.

\bibitem{chen2018sample}
Y.~Chen, Y.~Assael, B.~Shillingford, D.~Budden, S.~Reed, H.~Zen, Q.~Wang, L.~C.
  Cobo, A.~Trask, B.~Laurie \emph{et~al.}, ``Sample efficient adaptive
  text-to-speech,'' \emph{arXiv preprint arXiv:1809.10460}, 2018.

\bibitem{9054301}
H.~B. {Moss}, V.~{Aggarwal}, N.~{Prateek}, J.~{González}, and
  R.~{Barra-Chicote}, ``Boffin tts: Few-shot speaker adaptation by bayesian
  optimization,'' in \emph{ICASSP 2020 - 2020 IEEE International Conference on
  Acoustics, Speech and Signal Processing (ICASSP)}, 2020, pp. 7639--7643.

\bibitem{ren2019fastspeech}
Y.~Ren, Y.~Ruan, X.~Tan, T.~Qin, S.~Zhao, Z.~Zhao, and T.-Y. Liu, ``Fastspeech:
  Fast, robust and controllable text to speech,'' in \emph{NeurIPS 2019},
  November 2019.

\bibitem{Yu2019DurIANDI}
C.~Yu, H.~Lu, N.~Hu, M.~Yu, C.~Weng, K.~Xu, P.~Liu, D.~Tuo, S.~Kang, G.~Lei,
  D.~Su, and D.~Yu, ``Durian: Duration informed attention network for
  multimodal synthesis,'' \emph{ArXiv}, vol. abs/1909.01700, 2019.

\bibitem{Bahdanau2015NeuralMT}
D.~Bahdanau, K.~Cho, and Y.~Bengio, ``Neural machine translation by jointly
  learning to align and translate,'' \emph{CoRR}, vol. abs/1409.0473, 2015.

\bibitem{Kim2017ResidualLD}
J.~Kim, M.~El-Khamy, and J.~Lee, ``Residual lstm: Design of a deep recurrent
  architecture for distant speech recognition,'' in \emph{INTERSPEECH}, 2017.

\bibitem{Fan2015MultispeakerMA}
Y.~Fan, Y.~Qian, F.~K. Soong, and L.~He, ``Multi-speaker modeling and speaker
  adaptation for dnn-based tts synthesis,'' \emph{2015 IEEE International
  Conference on Acoustics, Speech and Signal Processing (ICASSP)}, pp.
  4475--4479, 2015.

\end{thebibliography}

\end{document}